\documentclass[12pt]{iopart}
\usepackage{iopams}
\usepackage{setstack}
\usepackage{graphicx}
\usepackage{epsfig}
\usepackage{tensor}
\usepackage[usenames]{color}

\newcommand{\sect}{\Sref}
\newcommand{\eqr}{\eref}
\newcommand{\g}{g\indices}

\usepackage[colorlinks=true,linkcolor=blue,citecolor=blue,urlcolor=blue]{hyperref}
\begin{document}

\title[L. Rodriguez and T. Yildirim]{Entropy and Temperature From Black-Hole/Near-Horizon-CFT Duality}

\author{Leo Rodriguez and Tuna Yildirim}

\address{Department of Physics and Astronomy, The University of Iowa, Iowa City, IA 52242}
\ead{\href{mailto:leo-rodriguez@uiowa.edu}{leo-rodriguez@uiowa.edu}}
\begin{abstract}
We construct a 2-dimensional CFT, in the form of a Liouville theory, in the near horizon limit of 4- and 3-dimensional black holes. The near horizon CFT assumes 2-dimensional black hole solutions first introduced by Christensen and Fulling \cite{chrisfull} and expanded to a greater class of black holes via Robinson and Wilczek \cite{robwill}. The 2-dimensional black holes admit a Diff($S^1$) subalgebra, which upon quantization in the horizon limit becomes Virasoro with calculable central charge. This charge and lowest Virasoro eigen-mode reproduce the correct Bekenstein-Hawking entropy of the 4- and 3-dimensional black holes via the known Cardy formula \cite{cardy2,cardy1}. Furthermore, the 2-dimensional CFT's energy momentum tensor is anomalous. However, In the horizon limit the energy momentum tensor becomes holomorphic equaling the Hawking flux of the 4- and 3-dimensional black holes. This encoding of both entropy and temperature provides a uniformity in the calculation of black hole thermodynamic and statistical quantities for the non local effective action approach.
\end{abstract}

\maketitle
\section{Introduction}\label{sec:intro}
The universally accepted thermodynamic and statistical properties of black holes have provided a unique insight and test for theories of quantum gravity. Though a fully formulated quantum field theory of gravity is lacking, a multitude of candidates exists, with string theory and loop quantum gravity leading in popularity. Despite a variety of theories, any serious candid should reproduce the correct form of the Bekenstein-Hawking entropy \cite{beken}
\begin{eqnarray}
\label{eq:bhe}
S_{BH}=&\frac{A}{4\hbar G}
\end{eqnarray}
and Hawking temperature \cite{hawk2,hawk3}
\begin{eqnarray}
\label{eq:ht}
T_H=&\frac{\hbar\kappa}{2\pi},
\end{eqnarray}
where $A$\footnote{$A=\int d^2x\sqrt{\left\vert\gamma\right\vert}$ where $\gamma_{ij}$ is obtained by setting $r=r_+$ for constant time. In the case for $Kerr$ $\sqrt{\left\vert\gamma\right\vert}=\left(r_+^2+J^2\right)\sin\theta$, which deviates from the standard form.} is the horizon area and $\kappa$ the surface gravity of the black hole. The fact that these quantities depend on both $\hbar$ and $G$ is evident of their quantum gravitational origin. For a comprehensive review of approaches to quantum gravity see \cite{au,carlip4,rov}.

Effective theories have had much success in reproducing \eqr{eq:bhe} and \eqr{eq:ht} via analysis of anomalous energy momentum tensors of non local effective actions \cite{wipf} and holographic 1- and 2-dimensional conformal field theories \cite{carlip4}, respectively. Yet, these methods remain in separate camps and a lack of uniformity in the derivation of both temperature and entropy in one simultaneous effective theory is missing. 

We will briefly review some of the pioneering work inspiring this note.
\subsection{Effective Action and Hawking Temperature}\label{subsec:efac}
Analysis of an anomalous energy momentum tensor to compute \eqr{eq:ht} was first carried out by Christensen and Fulling in \cite{chrisfull}. Considering the most general solution to the conservation equation
\begin{eqnarray}
\label{eq:stresscons}
\nabla_\mu T\indices{^\mu_\nu}=0,
\end{eqnarray}
they found that by restricting to the $r-t$ plane of a free scalar field in Schwarzschild geometry the energy momentum tensor exhibits a trace anomaly leading to the result:
\begin{eqnarray}
\label{}
\langle T\indices{^r_t}\rangle=\frac{1}{768\pi G^2 M^2}=\frac{\pi}{12}{T_H}^2,
\end{eqnarray}
which is exactly the luminosity (Hawking flux, Hawking radiation) of the 4-dimensional black hole in units $\hbar=1$.

A similar approach was studied in \cite{mukwipf} where the authors determined the $s$-wave contribution of a scalar field to the 4-dimensional effective action for an arbitrary spherically symmetric gravitational field. Applying their results to a Schwarzschild black hole, the authors showed the energy momentum tensor of the non local effective action to contain the Hawking Flux. Other closely related approaches for scalar fields and 2-dimensional theories include \cite{balfab2,balfab,cadtr,cadtr2,qpz}

Another method for computing Hawking Radiation first introduced by Robinson and Wilczek \cite{robwill} considers a quantum chiral-scalar field theory of 2-dimensions in the near horizon limit of a static 4-dimensional black hole. A 2-dimensional chiral field theory is know to exhibit a gravitational anomaly of the form:
\begin{eqnarray}
\label{eq:ga}
\nabla_\mu T\indices{^\mu_\nu}=\frac{1}{96\pi\sqrt{-g^{(2)}}}\epsilon^{\gamma\rho}\partial_\rho\partial_\lambda\Gamma\indices{^{(2)}^\lambda_\nu_\gamma},
\end{eqnarray}
where $g\indices{^{(2)}_\mu_\nu}$ contains the leftover components of the 4-dimensional metric which are not redshifted away in the near horizon limit of the functional
\begin{eqnarray}
\label{eq:scalar}
S_{\mbox{free scalar}}=\frac12\int d^4x\sqrt{-g}\nabla_\mu\varphi\nabla^\mu\varphi.
\end{eqnarray}
Robinson and Wilczek showed in the near horizon regime of a Schwarzschild black hole, that  to ensure a unitary quantum field theory the black hole should radiate as a thermal bath of temperature equaling $T_H$. In other words, quantum gravitational effects in the near horizon regime cancel the chiral/gravitational anomaly \cite{srv}. This method has been expanded to include gauge/gravitational anomalies and covariant anomalies \cite{rabin3,rabin,rabin2,rabin4} and has successfully reproduced the correct black hole temperature for charged-rotating black holes \cite{isowill,msoda}, $dS/AdS$ black holes \cite{gango,Jin}, rotating $dS/AdS$ black holes \cite{Jinwu,chen}, black rings and black string \cite{chen2,pwu}, 3-dimensional black holes \cite{nampark,setare} and black holes of non spherical topologies \cite{petro}. This method provides a fundamental reason for black hole thermodynamics based on symmetry principles of a near horizon quantum field theory. It also provides a 2-dimensional analogue for higher dimensional black holes besides the Schwarzschild case. This is a rather useful fact since the Ricci tensor in 2-dimensions is always Einstein with proportionality constant $\frac12\left(g\indices{^{(2)}^\mu^\nu}R\indices{^{(2)}_\mu_\nu}\right)$\footnote{In 2-dimensions the curvature of any Riemannian manifold is completely characterized by its scalar variant. This is because any $2$-form has only one independent component. Thus for any Riemannian-Levi-Cevitia connection $2$-form $\omega_{\alpha\beta}$, $d\omega_{12}=K\mbox{vol}^2$,where $K=\frac{1}{(2)((2)-1)}R^{(2)}$ is the Gauss curvature}. Thus classically, in 2-dimensions, there are no general relativistic dynamics and any gravitational effects that are present must have quantum gravitational origin with semi-classical metric $g\indices{^{(2)}_\mu_\nu}$.
\subsection{Holographic CFT and Entropy}\label{subsec:entro}
Applying the seminal work of Brown and Henneaux\footnote{The algebra of the asymptotic symmetry group of 3-dimensional gravity is Virasoro with calculable central charge.} \cite{brownhenau}, Strominger showed that the Bekenstein-Hawking entropy for $AdS_{3}$ black holes arises naturally from microscopic states of an asymptotic CFT via Cardy's formula \cite{strom2}. This work has been generalized to a class of rotating and $dS/AdS$ black holes in various dimensions in both near horizon regime and asymptotic infinity by Carlip and others \cite{carlip,carlip3,carlip2,kkp,silva,bgk}. Recently, Barnes, Vaman and Wu generalized the method to include 4-dimensional charged non-rotating black holes \cite{vaman} via uplifting the 4-dimensional black holes to 6-dimensional pure gravitational solutions. Though slightly different in their approaches, the general idea is as follows: there exists a set of diffeomorphisms preserving vector fields either in the near horizon/asymptotic symmetries of the metric or boundary conditions at spacial infinity as in \cite{silva}. This set of diffeomorphisms includes a Diff($S^1$) subalgebra parametrized by some discrete set of vector fields $\mathbf{\xi}_n$ for all $n$ $\in$ $\mathbb{Z}$ such that:
\begin{eqnarray}
\label{eq:diff}
i\left\{\mathbf{\xi}_m,\mathbf{\xi}_n\right\}=(m-n)\mathbf{\xi}_{m+n}
\end{eqnarray}
Upon quantization, Brown and Henneaux showed
\begin{eqnarray}
\label{eq:vir}
\left[\mathbf{\xi}_m,\mathbf{\xi}_n\right]=(m-n)\mathbf{\xi}_{m+n}+\frac{c}{12}m\left(m^2-1\right)\delta_{m+n,0}
\end{eqnarray}
where $c$ is a calculable central extension. The Bekenstein-Hawking Entropy is then given by Cardy's Formula \cite{cardy2,cardy1} in terms of $c$ and $\Delta_0$, where $\Delta_0$ is the eigen-mode of $\mathbf{\xi}_0$, via:
\begin{eqnarray}
\label{eq:cf}
S=2\pi\sqrt{\frac{c\cdot\Delta_0}{6}}
\end{eqnarray}
Applying the above outline to the 2-dimensional dilaton black hole
\begin{eqnarray}
\label{eq:2ddbh}
\eqalign{
ds^2&=g\indices{^D_\mu_\nu}dx^\mu dx^\nu\\
&=-\left[\left(\lambda x\right)^2-\frac{2M}{\lambda}\left(\lambda x\right)^3\right]dt^2+\left[\left(\lambda x\right)^2-\frac{2M}{\lambda}\left(\lambda x\right)^3\right]^{-1}dx^2,
}	
\end{eqnarray}
Cadoni computed the following central extension and zero mode \cite{cadss}:
\begin{eqnarray}
\label{eq:ccl0}
c=48\frac{M^2}{\lambda^2}~\mbox{and}~\Delta_0=\frac{M^2}{2\lambda^2}
\end{eqnarray}
 Cadoni further showed by conformally mapping \eqr{eq:2ddbh} to the $s$-wave sector of the Schwarzschild metric:
\begin{eqnarray}
\label{eq:ct1}
g\indices{^{(2)}_\mu_\nu}=2\phi g\indices{^{D}_\mu_\nu}
\end{eqnarray}
with
\begin{eqnarray}
\label{eq:ssct}
\lambda^2=\frac{1}{G}
\end{eqnarray}
and
\begin{eqnarray}
x=\frac{G}{r}
\end{eqnarray} 
\eqr{eq:ccl0} and \eqr{eq:cf} reproduced the correct Bekenstein-Hawking Entropy for the respective 4-dimensional black hole. In other words, together with Robinson and Wilczek's results \cite{robwill} both entropy and temperature of the Schwarzschild black hole induces some semi-classical theory of spacetieme
 \begin{eqnarray}
\label{eq:sssw}
ds^2=-\left(1-\frac{2GM}{r}\right)dt^2+\left(1-\frac{2GM}{r}\right)^{-1}dr^2.
\end{eqnarray}
Our goal will be to expound and extend this idea to include a greater class of rotating, charged and other types of black holes.
\subsection{$Kerr$/CFT Correspondence}\label{subsec:bh/cft}
A recent study by Guica, Hartman, Song and Strominger \cite{kerrcft} proposed that the near horizon geometry of an extremal Kerr black hole is holographically dual to a 2-dimensional chiral CFT\footnote{Distinct from the 2-dimensional chiral-scalar field theory employed by Robinson and Wilczek.} with non vanishing left central extension $c_L$. By constructing the Frolov-Thorne vacuum for generic Kerr geometry \cite{frolovthorne}, which reduces to the Hartle-Hawking vacuum \cite{hartlehawking} for $J\rightarrow0$, the authors obtain a non vanishing left Frolov-Thorne temperature $T_L$ in the near horizon, extremal limit. This temperature together with $c_L$ inside the Thermal Cardy Formula \cite{kerrcftsugra} reproduces the correct  Bekenstein-Hawking entropy for the extremal Kerr black hole:
\begin{eqnarray}
\label{eq:tcf}
S_{BH}=\frac{\pi^2}{3}c_LT_L
\end{eqnarray}
This correspondence has been extended to various exotic black holes in string theory, higher dimensional theories and gauged supergravities to name a few \cite{kerrcftstring,kerrcftsugra,kerrcftind}. 

One of the main arguments of the $Kerr$/CFT correspondence is to apply the rich ideas of holographic duality to more astrophysical objects/black-holes, such as the nearly extremal GRS 1915+105, a binary black hole system 11000$pc$ away in Aquila \cite{grs}. In \cite{kerrcft} the authors show that GRS 1915+105 is holographically dual to a 2-dimensional chiral CFT with $c_L=(2\pm1)\times 10^{79}$ and in the extremal limit the inner most stable circular orbit corresponds to the horizon. Thus, the authors conclude, any radiation emanating form the inner most circular orbit should be well described by the 2-dimensional chiral CFT, making the $Kerr$/CFT correspondence an essential theoretical tool in an astrophysical observation.

Despite the various models observationalists employ they all incorporate four main quantities: black hole mass $M$ and spin $J$, poloidal magnetic field at the horizon $B_0$ and (Eddington) luminosity $L$ for both supermassive \cite{bbhs} and stellar \cite{xpropbh} black holes. This provides a new testable playing field for holography, i.e. to use some induced 2-dimensional CFT in the near horizon regime of extremal and non-extremal black holes to model the four main quantities in accordance with observation. In particular, the origin or mechanism of $B_0$ is unclear from a theoretical standpoint, since it must be due to an accreting disk for non gauged black holes. Yet, it might find its origin in some black-hole/CFT duality.

The goal of this note is to model the near horizon regime via a 2-dimensional CFT as in the $Kerr$/CFT correspondence. This is done by combining the ideas from effective action approach and holographic duality and thus encodes both Hawking temperature and Bekenstein-Hawking entropy. Our final results hold in both extremal and non extremal cases.
\subsection{Outline of Main Idea}\label{subsec:outline}
We will model the near horizon regime with a 2-dimensional Liouville type quantum field theory, which is well understood \cite{seiberg},
\begin{eqnarray}
\label{eq:louiact}
S_{Liouville}=\frac{1}{96\pi}\int d^2x\sqrt{-g^{(2)}}\left\{-\Phi\square_{g^{(2)}}\Phi+2\Phi R^{(2)}\right\}.
\end{eqnarray}
We make this choice based on the fact that in this regime all mass and angular terms of \eqr{eq:scalar} fall of exponentially fast upon transformation from $r\rightarrow r_*$, were $\frac{\partial r}{\partial r_*}=f(r)$ \cite{robwill} and $g\indices{^{(2)}_t_t}=-f(r)$. This leaves us with an infinite collection of 2-dimensional free scalars in spherically symmetric spacetime $g\indices{^{(2)}_\mu_\nu}$\footnote{$g\indices{^{(2)}_\mu_\nu}$ may always be assumed  spherically symmetric since any Riemannian Space in 2-dimensions is conformally flat.}. The effective action of 2-dimensional free scalars is given by the Polyakov action \cite{wini,polyak}:
\begin{eqnarray}
\label{eq:polact}
\Gamma_{Polyakov}=\frac{1}{96\pi}\int d^2x\sqrt{-g^{(2)}}R^{(2)}\frac{1}{\square_{g^{(2)}}}R^{(2)}
\end{eqnarray}
and integrating out $\Phi$ in $S_{Liouville}$ yields $\Gamma_{Polyakov}$. In the case where the original 4-dimensional metric is not spherically symmetric \cite{isowill,msoda}, a $U(1)$ gauge sector appears in addition to \eqr{eq:polact}, which adds a gauge anomaly to Robinson and Wilczek's method for computing Hawking Radiation. Yet in our analysis, in addition to the Hawking flux, the presence of the $U(1)$ field only contributes a $\sim\left.\frac{e^2}{4\pi}\mathcal{A}^2_t\right\vert_{Horizon}$ to the holomorphic energy momentum tensor, which is the total flux of the gauge field. Thus it will suffice to only consider the field theory of \eqr{eq:louiact} since we are interested in Hawking effects.

The energy momentum tensor for \eqr{eq:louiact} is defined as:
\begin{eqnarray}
\label{eq:emt}
\eqalign{
\left\langle T_{\mu\nu}\right\rangle&=-\frac{2}{\sqrt{-g^{(2)}}}\frac{\delta S_{Liouville}}{\delta g\indices{^{(2)}^\mu^\nu}}\\
&=-\frac{1}{48\pi}\left\{\partial_\mu\Phi\partial_\nu\Phi-2\nabla_\mu\partial_\nu\Phi+g\indices{^{(2)}_\mu_\nu}\left[2R^{(2)}-\frac12\nabla_\alpha\Phi\nabla^\alpha\Phi\right]\right\}
}
\end{eqnarray}
and the equation of motion for the auxiliary scalar $\Phi$ is:
\begin{eqnarray}
\label{eq:eqmp}
\square_{g^{(2)}}\Phi=R^{(2)}
\end{eqnarray}
As an ansatz for the 2-dimensional metric $g\indices{^{(2)}_\mu_\nu}$, we choose the dimensional reduced spacetimes obtained via Robinson and Wilczek. Thus given a $g\indices{^{(2)}_\mu_\nu}$ we are free to solve \eqr{eq:eqmp} and \eqr{eq:emt} up to integration constants. Finally adopting Unruh Vacuum boundary conditions \cite{unruh}
\begin{eqnarray}
\label{eq:ubc}
\cases{
T_{++}=0&$r\rightarrow\infty,~l\rightarrow\infty$\cr
T_{--}=0&$r\rightarrow r_+$\cr
}
\end{eqnarray}
where $x^{\pm}=t\pm r_*$ are light-cone coordinates, $r_+$ is the horizon radius defined as the largest real root of $f(r)=0$ and $l$ is the de Sitter radius, all relevant integration constants are determined. At the horizon and asymptotic infinity for $\left(\Lambda=\pm\frac{1}{l^2}\right)=0$ and at the horizon only for $\Lambda\neq0$, \eqr{eq:emt} will be dominated by one holomorphic component. This component equals the Hawking flux of the 4- and 3-dimensional black holes, which determines the Hawking temperature.

The entropy will be determined by counting the horizon microstates of $g\indices{^{(2)}_\mu_\nu}$ via the Cardy formula \eqr{eq:cf}. Following the outline proposed in \cite{carlip2} we construct a near horizon Diff($S^1$) subalgebra satisfying \eqr{eq:diff} based on the isometries of $g\indices{^{(2)}_\mu_\nu}$. In the horizon limit ($\mathcal{I}^+$ boundary) the Diff($S^1$) subalgebra takes the form:
\begin{eqnarray}
\label{eq:diffppb}
i2\left\{\mathbf{\xi}^+_m,\mathbf{\xi}^+_n\right\}=(m-n)\mathbf{\xi}^+_{m+n},
\end{eqnarray}
where the factor 2 comes from neglecting the asymptotic infinity limit ($\mathcal{I}^-$ boundary). On the $\mathcal{I}^+$ boundary the energy momentum tensor is holomorphic given by the $\left\langle T_{++}\right\rangle$ component. Next, we define the charge on the $\mathcal{I}^+$ boundary\footnote{$Q_n$ is only conserved on the $\mathcal{I}^+$ boundary.}
\begin{eqnarray}
\label{eq:ccppb}
Q_n=\frac{3A}{\pi G}\int dx^+\left\langle T_{++}\right\rangle\mathbf{\xi}^+_n,
\end{eqnarray}
where $A$ is the horizon area of the higher dimensional black hole. The coefficient on the integral of $Q_n$ is chosen such that in the case when the higher dimensional black hole is Schwarzschild $Q_n=\frac{1}{16\pi G}\int dx^+\frac{12}{\pi \left(T^2_H\right)^2}\left\langle T_{++}\right\rangle\mathbf{\xi}^+_n$, where we have normalized the units of the energy momentum tensor. For a 1-dimensional CFT with holomorphic energy momentum tensor $T(z)$ we have \cite{cft}:
\begin{eqnarray}
\label{eq:emtt}
\delta_{\mathbf{\xi}(z)}T(z)=\mathbf{\xi} T'+2T\mathbf{\xi}'+\frac{k}{24\pi}\mathbf{\xi}''',
\end{eqnarray}
where $k$ is the central extension associated with the CFT \eqr{eq:louiact}. In the case for 2-dimensional quantum scalar, $k=1$ in agreement with \eqr{eq:tra}. Thus, given the transformation \eqr{eq:emtt} and compactifying the $\mathcal{I}^+$ boundary to a circle with period $\left(1/2\cdot1/T_H\right)$, where the $1/2$ takes the $\mathcal{I}^-$ boundary into account, we obtain the following charge algebra:
\begin{eqnarray}
\label{eq:ca}
\left[Q_m,Q_n\right]=(m-n)Q_n+\frac{c}{12}m\left(m^2-1\right)\delta_{m+n,0},
\end{eqnarray}
where $c$ is the central extension associated with the horizon-microstates of $g\indices{^{(2)}_\mu_\nu}$. The Bekenstein-Hawking entropy of the 4- and 3-dimensional black holes is then given by $Q_0$ and $c$ via \eqr{eq:cf}.

Finally we compare our results to \cite{cadss} by conformally mapping $g\indices{^{D}_\mu_\nu}$ to $g\indices{^{(2)}_\mu_\nu}$
\begin{eqnarray}
\label{eq:ct}
g\indices{^{(2)}_\mu_\nu}=2\phi g\indices{^{D}_\mu_\nu}
\end{eqnarray}
for some conformal factor $2\phi=(\lambda x)^{-2}$, where \eqr{eq:ccl0} is assumed invariant under conformal transformations \cite{cadtr2,cadmigceq}. We should not that the dimensionally reduced spacetimes satisfy the equations of motion for 2-dimesnional dilaton gravity in the Schwarzschild case only. This is impart due to the fact that 2-dimensional dilaton gravity is a dimensionally reduced theory of the pure Einstin-Hilbert action with metric ansatz:
\begin{eqnarray}
\label{eq:dma}
ds_{(4)}^2=ds^2_{(2)}+\frac{2}{\lambda^2}\phi d\Omega^2_{(2)},
\end{eqnarray}
where $ds^2_{(2)}$ is as in \eqr{eq:2ddbh}, with no additional matter sources. Now, since the above metric ansatz does not incorporate axisymmetric solutions and the Reissner-Nordstr\"om solves Einstein-Hilbert-Maxwell theory, it is clear why their 2-dimensional counterparts do not solve the same field equations. Despite this short coming, we conjecture that the conformal map \eqr{eq:ct} still encodes the correct microstates for $g\indices{^{(2)}_\mu_\nu}$ based on comparisons of respective entropies obtained from \eqr{eq:ca} and \eqr{eq:ct} applied to the Reissner-Nordstr\"om case. This may be impart, since in 2-dimensions there are no classical general relativistic dynamics, as mentioned in the \hyperref[sec:intro]{Introduction}, and $\g{^{(2)}_\mu_\nu}$ may be a dimensinally reduced effective metric of a ultraviolet complete 4-dimensional theory of gravity.
\section{Examples from 4- and 3-Dimensions}\label{sec:4d}
We will now apply the method, outlined above, to  various 4- and 3-dimensional black holes with zero and non zero cosmological constant and construct their Hawking flux, associated entropy and temperature.
\subsection{Spherically Symmetric Solutions}\label{subsec:sphs}
In this class we will consider the Scwarzschild ($SS$) and Reissner-Nordstr\"om ($RNS$) black holes. Their 2-dimensional analogues have the form \cite{isowill2,robwill}
\begin{eqnarray}
\label{eq:ssrns2d}
g\indices{^{(2)}_\mu_\nu}=
\left(\begin{array}{cc}-f(r) & 0 \\0 & \frac{1}{f(r)}\end{array}\right),
\end{eqnarray}
where
\begin{eqnarray}
\label{eq:fss}
f_{SS}(r)=1-\frac{2GM}{r}
\end{eqnarray}
and 
\begin{eqnarray}
\label{eq:frns}
f_{RNS}(r)=1-\frac{2GM}{r}+\frac{Q^2G}{r^2}
\end{eqnarray}
Next, using the above ansatz and solving \eqr{eq:eqmp} we get:
\begin{eqnarray}
\label{eq:pss}
\Phi_{SS}=C_2t+C_1r+\ln{r}-\left(1 - 2GMC_1\right)\ln{\left(r - 2 G M\right)}+C_3
\end{eqnarray}
and 
\begin{eqnarray}
\label{eq:prns}
\eqalign{
\Phi_{RNS}=&C_2t+C_1r+\frac{C_1\sqrt{G}\left(2GM^2-Q^2\right)}{\sqrt{GM^2-Q^2}}\arctan{\left(\frac{GM-r}{\sqrt{G^2M^2-GQ^2}}\right)}\\
&+2\ln{r}-\left(1 - GMC_1\right)\ln{\left(r^2 - 2GMr+GQ^2\right)}+C_3
}
\end{eqnarray}
Using these auxiliary fields in \eqr{eq:emt} and transforming to light cone coordinates we obtain:
\begin{eqnarray}
\label{eq:tss}
\eqalign{
\left\langle T^{SS}_{++}\right\rangle=\frac{-r^4 (C_1+C_2)^2+8rGM-12G^2M^2}{192 \pi r^4}\\
\left\langle T^{SS}_{--}\right\rangle=\frac{-r^4 (C_1-C_2 )^2+8rG M-12G^2M^2}{192 \pi r^4}\\
\left\langle T^{SS}_{+-}\right\rangle=\left\langle T^{SS}_{-+}\right\rangle=\frac{G M (r-2 G M)}{24 \pi r^4}
}
\end{eqnarray}
and 
\begin{eqnarray}
\label{eq:trns}
\eqalign{
\left\langle T^{RNS}_{++}\right\rangle=&\left[-r^6 (C_1+C_2 )^2+8 r^3 G M-12 r^2 G \left(G M^2+Q^2\right)\right.\\
&\left.+24 rG^2 M Q^2-8 G^2 Q^4\right]/\left[192 \pi  a^6\right]\\
\left\langle T^{RNS}_{--}\right\rangle=&\left[-r^6 (C_1-C_2 )^2+8 r^3 G M-12 r^2 G \left(G M^2+Q^2\right)\right.\\
&\left.+24 rG^2 M Q^2-8 G^2 Q^4\right]/\left[192 \pi  a^6\right]\\
\left\langle T^{RNS}_{+-}\right\rangle=&\left\langle T^{RNS}_{-+}\right\rangle=\frac{G \left(2 r M-3 Q^2\right) \left(r^2-2 r G M+G Q^2\right)}{48 \pi 
   r^6}
}
\end{eqnarray}
The fact that both energy momentum tensors are not holomorphic/anti-homolomorphic signals the existence of a conformal anomaly taking the form
\begin{eqnarray}
\label{eq:tra}
\left\langle T\indices{_\mu^\mu}\right\rangle=-\frac{1}{24\pi}R^{(2)}
\end{eqnarray}
due to the trace anomaly \cite{cft}. Imposing \eqr{eq:ubc} to eliminate $C_1$ and $C_2$ our final steps are to analyze $\left\langle T\indices{_\mu_\nu}\right\rangle$ at the horizon and construct the conformal map \eqr{eq:ct}. At the horizon the energy momentum tensors are dominated by one holomorphic component $\left\langle T_{++}\right\rangle$ given by
\begin{eqnarray}
\label{eq:hfss}
\left\langle T_{++}^{SS}\right\rangle=\frac{1}{768 \pi  G^2 M^2}=\frac{\pi}{12}\left(T_H\right)^2
\end{eqnarray}
and 
\begin{eqnarray}
\label{eq:hfrns}
\eqalign{
\left\langle T_{++}^{RNS}\right\rangle=&\frac{G^2 \left(G M^2-Q^2\right) \left(2 M \sqrt{G \left(G
   M^2-Q^2\right)}+2 G M^2-Q^2\right)}{48 \pi  \left(\sqrt{G \left(G
   M^2-Q^2\right)}+G M\right)^6}\\
   &=\frac{\pi}{12}\left(T_H\right)^2,
}
\end{eqnarray}
which are in agreement with Hawking's original results \cite{hawk2,hawk3}.

Following \cite{carlip2}, we compute the near horizon diffeomorphisms satisfying \eqr{eq:diff}. We get:
\begin{eqnarray}
\label{eq:diffss}
\mathbf{\xi}^{SS}_n=4 G M e^{i n \kappa t}\partial_t+\frac{i n (r-2 G M)^2 e^{i n \kappa t}}{G M}\partial_r
\end{eqnarray}
and
\begin{eqnarray}
\label{eq:diffrns}
\eqalign{
\mathbf{\xi}^{RNS}_n=&\frac{\left(\sqrt{G \left(G M^2-Q^2\right)}+G M\right)^3 e^{i \kappa n t}}{G
   \left(M \sqrt{G \left(G M^2-Q^2\right)}+G M^2-Q^2\right)}\partial_t+\\
   &\frac{i n e^{i \kappa n t} \left(r^2-2 r G M+G Q^2\right)^2}{r G \left(r
   M-Q^2\right)}\partial_r,
}
\end{eqnarray}
where $\kappa$ is the horizon surface gravity. Transforming to $x^\pm$ coordinates and taking the horizon limit, we obtain the $\mathcal{I}^+$ boundary charge algebras:
\begin{eqnarray}
\label{eq:cass}
\left[Q_m,Q_n\right]_{SS}=(m-n)Q_n+2GM^2m\left(m^2-1\right)\delta_{m+n,0},
\end{eqnarray}
and 
\begin{eqnarray}
\label{eq:carns}
\eqalign{
\left[Q_m,Q_n\right]_{RNS}=&(m-n)Q_n+\\
&\left(M \sqrt{G \left(G M^2-Q^2\right)}+G M^2-\frac{Q^2}{2}\right)m\left(m^2-1\right)\delta_{m+n,0},
}
\end{eqnarray}
which imply
\begin{eqnarray}
\label{eq:caczss}
Q^{SS}_0=&GM^2\\
c^{SS}=&24GM^2
\end{eqnarray}
and 
\begin{eqnarray}
\label{eq:caczrns}
Q^{RNS}_0=&\frac{\left(\sqrt{G \left(G M^2-Q^2\right)}+G M\right)^2}{4 G}\\
c^{RNS}=&\frac{6 \left(\sqrt{G^2 M^2-G Q^2}+G M\right)^2}{G}
\end{eqnarray}
and using these values in \eqr{eq:cf} we get:
\begin{eqnarray}
\label{eq:caentss}
S_{SS}=4\pi GM^2=\frac{A}{4G}
\end{eqnarray}
and 
\begin{eqnarray}
\label{eq:caentrns}
S_{RNS}=\frac{\pi  \left(\sqrt{G \left(G M^2-Q^2\right)}+G M\right)^2}{G}=\frac{A}{4G}.
\end{eqnarray}

Finally, conformally mapping $g\indices{^{(2)}_\mu_\nu}$ to $g\indices{^{(D)}_\mu_\nu}$ implies:
\begin{eqnarray}
\label{eq:lxclss}
\lambda_{SS}^2=&\frac{1}{G}\\
x_{SS}=&\frac{G}{r}\\
c_{SS}=&48 G M^2\\
\Delta_{0}^{SS}=&\frac{G M^2}{2}
\end{eqnarray}
as in \cite{cadss} and 
\begin{eqnarray}
\label{eq:lxclrns}
\lambda_{RNS}^2=&\frac{4 G M^2}{\left(\sqrt{G^2 M^2-G Q^2}+G M\right)^2}\\
x_{RNS}=&\frac{G \left(2 r M-Q^2\right)}{2 r^2 M}\\
c_{RNS}=&\frac{12 \left(\sqrt{G^2 M^2-G Q^2}+G M\right)^2}{G}\\
\Delta_{0}^{RNS}=&\frac{\left(\sqrt{G^2 M^2-G Q^2}+G M\right)^2}{8 G}
\end{eqnarray}
where $\lambda_{RNS}$ and $x_{RNS}$ are such that
\begin{eqnarray}
\label{eq:constlx}
\eqalign{
\lim_{Q\to0}\lambda_{RNS}=\lambda_{SS}~\mbox{and}~\lim_{Q\to0}x_{RNS}=x_{SS}
}
\end{eqnarray}
Using \eqr{eq:cf} we find the respective entropies:
\begin{eqnarray}
\label{eq:entss}
S_{SS}=4\pi GM^2=\frac{A}{4G}
\end{eqnarray}
and 
\begin{eqnarray}
\label{eq:entrns}
S_{RNS}=\frac{\pi  \left(\sqrt{G \left(G M^2-Q^2\right)}+G M\right)^2}{G}=\frac{A}{4G}.
\end{eqnarray}
We see that our central extension and zero-mode relate to Cadoni's via
\begin{eqnarray}
c=\frac{c_c}{2}
\end{eqnarray}
and
\begin{eqnarray}
Q_0=&2\Delta_0.
\end{eqnarray}
Yet, their respective products are equal and produce entropies in agreement with the Bekenstein-Hawking area law \cite{beken} for $\hbar=1$. Thus in accord with our conjecture at the end of \sect{subsec:outline}, we choose to conformally map into Cadoni's solution for all $g\indices{^{(2)}_\mu_\nu}$ for calculational simplicity.

We will proceed to solidify our main argument by applying the methods of \sect{subsec:outline} to several more black hole solutions of various types.
\subsection{Axisymmetric Solutions}\label{subsec:axi}
For this class we analyze the Kerr ($K$) and Kerr-Newman ($KN$) Black holes with 2-dimensional analogues \cite{isowill,isowill2}
\begin{eqnarray}
\label{eq:kkn2d}
g\indices{^{(2)}_\mu_\nu}=
\left(\begin{array}{cc}-f(r) & 0 \\0 & \frac{1}{f(r)}\end{array}\right),
\end{eqnarray}
where
\begin{eqnarray}
\label{eq:fkkn}
f(r)=\frac{\Delta }{r^2+J^2}
\end{eqnarray}
and 
\begin{eqnarray}
\label{eq:dkkn}
\Delta=
\cases{
r^2-2 r G M+J^2&$K$\cr
r^2-2 r G M+G Q^2+J^2&$KN$\cr
}.
\end{eqnarray}
The auxiliary scalars read:
\begin{eqnarray}
\label{eq:ask}
\eqalign{
\Phi_{K}=&r C_1+t C_2+(C_1 G M-1) \log \left(r^2-2 r G M+J^2\right)+\log
   \left(r^2+J^2\right)+\\
   &\frac{2 C_1 G^2 M^2 \arctan\left(\frac{r-G
   M}{\sqrt{J^2-G^2 M^2}}\right)}{\sqrt{J^2-G^2 M^2}}+C_3
}
\end{eqnarray}
and
\begin{eqnarray}
\label{eq:askn}
\eqalign{
\Phi_{KN}=r A+tC_2+(C_1 G M-1) \log \left(r^2-2 r G M+G Q^2+J^2\right)+\\
\log\left(r^2+J^2\right)+\frac{C_1 G \left(2 G M^2-Q^2\right) \arctan\left(\frac{r-G M}{\sqrt{G \left(Q^2-GM^2\right)+J^2}}\right)}{\sqrt{G \left(Q^2-G M^2\right)+J^2}}+C_3
}
\end{eqnarray}
from which we obtain the energy momentum tensors:
\begin{eqnarray}
\label{eq:tk}
\eqalign{
\left\langle T^{K}_{++}\right\rangle=&-\left[r^8 (C_1+C_2 )^2+4 r^6 J^2 (C_1+C_2 )^2-8 r^5 G M+6 r^4
   \left(C_1^2 J^4+\right.\right.\\
   &\left.\left.2 C_1 J^4 C_2 +2 G^2 M^2+J^4 C_2 ^2\right)+16 r^3 GJ^2 M+4 r^2 \left(C_1^2 J^6+\right.\right.\\
   &\left.\left.2 C_1 J^6 C_2 -10 G^2 J^2 M^2+J^6 C_2^2\right)+24 r G J^4 M+C_2^2 J^8+\right.\\
   &\left.2 C_1 J^8 C_2 -4 G^2 J^4 M^2+J^8C_2 ^2\right]/\left[192 \pi  \left(r^2+J^2\right)^4\right]\\
\left\langle T^{K}_{--}\right\rangle=&-\left[r^8 (C_1-C_2 )^2+4 r^6 J^2 (C_1-C_2 )^2-8 r^5 G M+6 r^4
   \left(C_1^2 J^4-\right.\right.\\
   &\left.\left.2 C_1 J^4 C_2 +2 G^2 M^2+J^4 C_2 ^2\right)+16 r^3 GJ^2 M+4 r^2 \left(C_1^2 J^6-\right.\right.\\
   &\left.\left.2 C_1 J^6 C_2 -10 G^2 J^2 M^2+J^6 C_2^2\right)+24 r G J^4 M+C_1^2 J^8-\right.\\
   &\left.2 C_1 J^8 C_2 -4 G^2 J^4 M^2+J^8C_2 ^2\right]/\left[192 \pi  \left(r^2+J^2\right)^4\right]\\
\left\langle T^{K}_{+-}\right\rangle=&\left\langle T^{K}_{-+}\right\rangle=\frac{r G M \left(r^2-3 J^2\right) \left(r^2-2 r G M+J^2\right)}{24 \pi 
   \left(r^2+J^2\right)^4}
}
\end{eqnarray}
and 
\begin{eqnarray}
\label{eq:tkn}
\eqalign{
\left\langle T^{KN}_{++}\right\rangle=&-\left[r^8 (C_1+C_2 )^2+4 r^6 J^2 (C_1+C_2 )^2-8 r^5 G M+6 r^4
   \left(C_1^2 J^4+\right.\right.\\
   &\left.\left.2 C_1 J^4 C_2 +2 G^2 M^2+2 G Q^2+J^4 C_2 ^2\right)-8r^3 G M \left(3 G Q^2-\right.\right.\\
   &\left.\left.2 J^2\right)+4 r^2 \left(C_1^2 J^6+2 C_1 J^6 C_2+2 G^2 \left(Q^4-5 J^2 M^2\right)+\right.\right.\\
   &\left.\left.2 G J^2 Q^2+J^6 C_2 ^2\right)+24r G J^2 M \left(G Q^2+J^2\right)+C_1^2 J^8+\right.\\
   &\left.2 C_1 J^8 C_2 -4 G^2 J^4
   M^2-4 G^2 J^2 Q^4-4 G J^4 Q^2+J^8 C_2 ^2\right]/\\
   &\left[192 \pi\left(r^2+J^2\right)^4\right]\\
\left\langle T^{KN}_{--}\right\rangle=&-\left[r^8 (C_1-C_2 )^2+4 r^6 J^2 (C_1-C_2 )^2-8 r^5 G M+6 r^4
   \left(C_1^2 J^4-\right.\right.\\
   &\left.\left.2 C_1 J^4 C_2 +2 G^2 M^2+2 G Q^2+J^4 C_2 ^2\right)-8r^3 G M \left(3 G Q^2-\right.\right.\\
   &\left.\left.2 J^2\right)+4 r^2 \left(C_1^2 J^6-2 C_1 J^6 C_2+2 G^2 \left(Q^4-5 J^2 M^2\right)+\right.\right.\\
   &\left.\left.2 G J^2 Q^2+J^6 C_2 ^2\right)+24r G J^2 M \left(G Q^2+J^2\right)+C_1^2 J^8-\right.\\
   &\left.2 C_1 J^8 C_2 -4 G^2 J^4M^2-4 G^2 J^2 Q^4-4 G J^4 Q^2+J^8 C_2 ^2\right]/\\
   &\left[192 \pi 
   \left(r^2+J^2\right)^4\right]\\
\left\langle T^{KN}_{+-}\right\rangle=&\left\langle T^{KN}_{-+}\right\rangle=\left[G \left(2 r^3 M-3 r^2 Q^2-6 r J^2 M+J^2 Q^2\right) \left(r^2-\right.\right.\\
   &\left.\left.2 r G
   M+G Q^2+J^2\right)\right]/\left[48 \pi  \left(r^2+J^2\right)^4\right]
}
\end{eqnarray}
which exhibits conformal anomaly \eqr{eq:tra}. Applying \eqr{eq:ubc} and taking the horizon limit we obtain the holomorphic pieces
\begin{eqnarray}
\label{eq:hfk}
\left\langle T_{++}^{K}\right\rangle=\frac{\pi  \left(G^2 M^2-J^2\right)}{12 \left(4 \pi  G M \sqrt{G^2
   M^2-J^2}+4 \pi  G^2 M^2\right)^2}=\frac{\pi}{12}\left(T_H\right)^2
\end{eqnarray}
and 
\begin{eqnarray}
\label{eq:hfkn}
\eqalign{
\left\langle T_{++}^{KN}\right\rangle=&-\frac{G \left(Q^2-G M^2\right)+J^2}{48 \pi  \left(\left(\sqrt{G^2 M^2-G
   Q^2-J^2}+G M\right)^2+J^2\right)^2}\\
   &=\frac{\pi}{12}\left(T_H\right)^2
   }
\end{eqnarray}
agreeing with Hawking's result \cite{hawk2,hawk3}. Next, from \eqr{eq:ct} and applying similar boundary conditions as in \eqr{eq:constlx} we obtain
\begin{eqnarray}
\label{eq:lxclk}
\lambda_{K}^2=&\frac{4 G M^2}{\left(\sqrt{G^2 M^2-J^2}+G M\right)^2+J^2}\\
x_{K}=&\frac{r G}{r^2+J^2}\\
c_{K}=&\frac{12 \left(\left(\sqrt{G^2 M^2-J^2}+G M\right)^2+J^2\right)}{G}\\
\Delta_{0}^{K}=&\frac{\left(\sqrt{G^2 M^2-J^2}+G M\right)^2+J^2}{8 G}
\end{eqnarray}
and 
\begin{eqnarray}
\label{eq:lxclkn}
\lambda_{KN}^2=&\frac{4 G M^2}{\left(\sqrt{G^2 M^2-G Q^2-J^2}+G M\right)^2+J^2}\\
x_{KN}=&\frac{2 a G M-G Q^2}{2 a^2 M+2 J^2 M}\\
c_{KN}=&\frac{12 \left(\left(\sqrt{G^2 M^2-G Q^2-J^2}+G M\right)^2+J^2\right)}{G}\\
\Delta_{0}^{KN}=&\frac{\left(\sqrt{G^2 M^2-G Q^2-J^2}+G M\right)^2+J^2}{8 G}
\end{eqnarray}
which give the respective entropies:
\begin{eqnarray}
\label{eq:entk}
S_{K}=2 \pi  M \left(\sqrt{G^2 M^2-J^2}+G M\right)=\frac{A}{4G}
\end{eqnarray}
and 
\begin{eqnarray}
\label{eq:entkn}
S_{KN}=\pi  \left(2 M \left(\sqrt{G \left(G M^2-Q^2\right)-J^2}+G
   M\right)-Q^2\right)=\frac{A}{4G}
\end{eqnarray}
reproducing the Bekenstein-Hawking area law \cite{beken} and continuing the trend of \sect{subsec:sphs}.
\subsection{Spherically Symmetric $SSdS$ and Rotating $BTZ$}\label{subsec:sitter}
Now, we turn our attention to black holes with non zero cosmological constant:
\begin{eqnarray}
\label{eq:cc}
\Lambda=
\cases{
\frac{1}{l^2}&$dS$\cr
-\frac{1}{l^2}&$AdS$\cr
}
,
\end{eqnarray}
where $l$ is the de Sitter radius. In this black hole class we consider the spherically symmetric $dS$ ($SSdS$) with line element
\begin{eqnarray}
\label{eq:ssds}
\eqalign{
ds^2=&-\left(1-\frac{2 G M}{r}-\frac{r^2 \Lambda }{3}\right)dt^2+\left(1-\frac{2 G M}{r}-\frac{r^2 \Lambda }{3}\right)^{-1}dr^2\\
&+r^2d\Omega
}
\end{eqnarray}
and the 3-dimensional $BTZ$ black hole with line element
\begin{eqnarray}
\label{eq:btz}
\eqalign{
ds^2=&-\left(-8GM+\frac{r^2}{l^2}+\frac{16GJ^2 }{r^2}\right)dt^2+\left(-8GM+\frac{r^2}{l^2}+\frac{16GJ^2 }{r^2}\right)^{-1}dr^2\\
&+r^2\left(d\phi-\frac{4GJ}{r^2}dt\right)^2.
}
\end{eqnarray}
Their 2-dimensional analogues \cite{setare,gango} are as in \eqr{eq:kkn2d} where
\begin{eqnarray}
\label{eq:fssdsbtz}
f(r)=
\cases{
1-\frac{2 G M}{r}-\frac{r^2 \Lambda }{3}&$SSdS$\cr
-8GM+\frac{r^2}{l^2}+\frac{16GJ^2 }{r^2}&$BTZ$\cr
}
\end{eqnarray}
Following the steps outlined in \sect{subsec:outline} we obtain the energy momentum tensors:
\begin{eqnarray}
\label{eq:tssds}
\eqalign{
\left\langle T^{SSdS}_{++}\right\rangle=&-\left[r^4 \left(3 C_1^2+6 C_1 C_2 +3 C_2 ^2-4 \Lambda \right)+24 r^3 G
   M \Lambda\right.\\
   &\left. -24 r G M+36 G^2 M^2\right]/\left[576 \pi  r^4\right]\\
\left\langle T^{SSdS}_{--}\right\rangle=&\left[r^4 \left(-3 C_1^2+6 C_1 C_2 -3 C_2 ^2+4 \Lambda \right)-24 r^3 G
   M \Lambda\right.\\
   &\left. +24 r G M-36 G^2 M^2\right]/\left[576 \pi  r^4\right]\\
\left\langle T^{SSdS}_{+-}\right\rangle=&\left\langle T^{SSdS}_{-+}\right\rangle=-\left[r^6 \Lambda ^2-3 r^4 \Lambda +12 r^3 G M \Lambda -18 r G M\right.\\
&\left.+36 G^2M^2\right]/\left[432 \pi  a^4\right]
}
\end{eqnarray}
and 
\begin{eqnarray}
\label{eq:tbtz}
\eqalign{
\left\langle T^{BTZ}_{++}\right\rangle=&-\left[r^6 \left(C_1^2 l^2+2 C_1 l^2 C_2 -32 G M+l^2 C_2 ^2\right)+384
   r^4 G^2 J^2\right.\\
   &\left.-1536 r^2 G^3 J^2 l^2 M+2048 G^4 J^4 l^2\right]/\left[192 \pi  r^6 l^2\right]\\
\left\langle T^{BTZ}_{--}\right\rangle=&-\left[r^6 \left(C_1^2 l^2-2 C_1 l^2 C_2 -32 G M+l^2 C_2 ^2\right)+384
   r^4 G^2 J^2\right.\\
   &\left.-1536 r^2 G^3 J^2 l^2 M+2048 G^4 J^4 l^2\right]/\left[192 \pi  r^6 l^2\right]\\
\left\langle T^{BTZ}_{+-}\right\rangle=&\left\langle T^{BTZ}_{-+}\right\rangle=-\left[\left(r^4+48 G^2 J^2 l^2\right) \left(r^4-8 r^2 G l^2 M\right.\right.\\
&\left.\left.+16 G^2 J^2l^2\right)\right]/\left[48 \pi  r^6 l^4\right]
}
\end{eqnarray}
with conformal anomaly \eqr{eq:tra}. Applying \eqr{eq:ubc} we obtain the holomorphic piece
\begin{eqnarray}
\label{eq:hfssds}
\left\langle T_{++}\right\rangle=&\frac{\pi}{12}\left(T_H\right)^2
\end{eqnarray}
for both spacetimes in their respective horizon limits and agreeing as before with Hawking's results \cite{hawk2,hawk3}. Next, their respective entropies are computed via \eqr{eq:cf}, \eqr{eq:ct},
\begin{eqnarray}
\label{eq:lxclssds}
\lambda_{SSdS}^2=&-\left[16 G M^2 \Lambda ^2 \left(\sqrt{\Lambda ^3 \left(9 G^2 M^2 \Lambda
   -1\right)}-3 G M \Lambda^2\right)^{2/3}\right]/\nonumber\\
   &\left[\left(\left(\sqrt{3}-i\right) \left(\sqrt{\Lambda ^3
   \left(9 G^2 M^2 \Lambda -1\right)}-3 G M \Lambda
   ^2\right)^{2/3}\right.\right.\\
   &\left.\left.+\left(-\sqrt{3}-i\right) \Lambda \right)^2\right]\nonumber\\
x_{SSdS}=&\frac{r^3 \Lambda +6 G M}{6 r M}\\
c_{SSdS}=&-\left[3 \left(\left(\sqrt{3}-i\right) \left(\sqrt{\Lambda ^3 \left(9 G^2
   M^2 \Lambda -1\right)}-3 G M \Lambda^2\right)^{2/3}\right.\right.\nonumber\\
   &\left.\left.+\left(-\sqrt{3}-i\right) \Lambda \right)^2\right]/\left[G \Lambda
   ^2 \left(\sqrt{\Lambda ^3 \left(9 G^2 M^2 \Lambda -1\right)}\right.\right.\\
   &\left.\left.-3 G M\Lambda ^2\right)^{2/3}\right]\nonumber\\
\Delta_{0}^{SSdS}=&-\left[\left(\left(\sqrt{3}-i\right) \left(\sqrt{\Lambda ^3 \left(9 G^2M^2 \Lambda -1\right)}-3 G M \Lambda
   ^2\right)^{2/3}\right.\right.\nonumber\\
   &\left.\left.+\left(-\sqrt{3}-i\right) \Lambda \right)^2\right]/\left[32 G\Lambda ^2 \left(\sqrt{\Lambda ^3 \left(9 G^2 M^2 \Lambda -1\right)}\right.\right.\\
   &\left.\left.-3G M \Lambda ^2\right)^{2/3}\right]\nonumber
\end{eqnarray}
and
\begin{eqnarray}
\label{eq:lxclbtz}
\lambda_{BTZ}^2=&\frac{4 G M^2}{\sqrt{\sqrt{G^2 l^2 \left(l^2 M^2-J^2\right)}+G l^2 M}}\\
x_{BTZ}=&\frac{(-r^4 + r^2 l^2 - 16 G^2 J^2 l^2 + 8 r^2 G l^2 M)}{(2 r^2 l^2 M)}\\
c_{BTZ}=&\frac{12 \sqrt{\sqrt{G^2 l^2 \left(l^2 M^2-J^2\right)}+G l^2 M}}{G}\\
\Delta_{0}^{BTZ}=&\frac{\sqrt{\sqrt{G^2 l^2 \left(l^2 M^2-J^2\right)}+G l^2 M}}{8 G}
\end{eqnarray}
reproducing the Bekenstein-Hawking area law \cite{beken} 
\begin{eqnarray}
\label{eq:entssds}
S=\frac{A}{4G}
\end{eqnarray}
in both cases via \eqr{eq:cf}. Thus by modeling the near horizon regime with a 2-dimensional CFT, we have computed both entropy and temperature of 4-dimensional Schwarzschild, Reissner-Nordstr\"om, Kerr, Kerr-Newman, Spherically Symmetric $dS$ and of 3-dimensional $BTZ$ black hole. 
\section{Conclusion}\label{sec:conclusion}
To conclude, we have analyzed quantum black hole properties via non local effective action in the near horizon regime. For a relatively large class of black holes, including $dS$ and $AdS$ solutions in 3- and 4-dimensions, both entropy and temperature are computed from 2-dimensional conformal field theory techniques in the near horizon regime. The 2-dimensional CFT was modeled via a Liouville action with 2-dimensional black hole solutions given by Robinson and Wilczek's dimensional reduction first discussed in \cite{robwill}. These 2-dimensional black holes exhibit a Diff($S^1$) subalgebra, up to conformal transformation, first discovered by Cadoni \cite{cadss} for the $s$-wave sector of a Schwarzschild black hole. Analysis of the anomalous energy momentum tensor of the Liouville theory and the Diff($S^1$) subalgebra reproduces the Hawking temperature and Bekenstein-Hawking entropy for the respective 4- and 3-dimensional black holes.  

The anomalous contribution \eqr{eq:tra} signals interesting physics as Christensen and Fulling showed \cite{chrisfull}. In this case the anomaly relates to quantum black hole physics and in fact for Schwarzschild 
\begin{eqnarray}
\label{eq:sstra}
\frac{1}{16}\left\langle T\indices{_\mu^\mu}\right\rangle=-\frac{\pi}{12}\left(T_H\right)^2,
\end{eqnarray}
which is not the case for any other spacetime considered above except in their respective holomorphic limits. Though the 2-dimensional trace anomaly did not factor much into the main calculations of this note, it remains an interesting feature to interpret especially for the non Schwarzschild cases. 

The methods outlined in \sect{subsec:outline} seem to be universal at least in 3- and 4-dimensions. It remains interesting, for future work, to generalize \eqr{eq:ccppb} and \eqr{eq:ca} to general black holes/strings in arbitrary dimensions as Peng, Wu  \cite{pwu} and Xu, Chen \cite{chen} have done for Robinson and Wilczek's gauge gravitational anomaly cancellation method. 
\ack{
We thank Vincent Rodgers for suggesting this problem and support. We thank Steven Carlip for enlightening discussions, comments and encouragement. We thank Edwin Barnes, Philip Kaaret, Diana Vaman and Leopoldo Pando Zayas for enlightening discussions. We also thank the entire \href{http://www-hep.physics.uiowa.edu/~vincent/Research/index.html}{Diffeomorphisms and Geometry Research Group} of the University of Iowa for continued support, helpful discussions and comments. 

We would also like to thank the editorial staff of \CQG and the referees for helping us correct our work and pointing us to interesting and useful literature.}

\vspace{.5cm}
\begin{center}
\noindent\line(1,0){150}
\end{center}
\bibliographystyle{unsrt}
\bibliography{cftgr}

\end{document}